**Article type:** Research Article

# Detecting collagen by machine learning improved photoacoustic spectral analysis for breast cancer diagnostics: feasibility studies with murine models

Jiayan Li[1], Lu Bai[2], Yingna Chen[1], Junmei Cao[1], Jingtao Zhu[3], Wenxiang Zhi[2,*], Qian Cheng[1,4,5,*]

[1] Institute of Acoustics, School of Physics Science and Engineering, Tongji University, Shanghai, P. R. China

[2] Department of Ultrasonography, Fudan University Shanghai Cancer Center, Shanghai Medical College, Fudan University, Shanghai, P. R. China

[3] School of Physics Science and Engineering, Tongji University, Shanghai, P. R. China

[4] National Key Laboratory of Autonomous Intelligent Unmanned Systems, P. R. China

[5] Frontiers Science Center for Intelligent Autonomous Systems, Ministry of Education, P. R. China

*Correspondence

Wenxiang Zhi, Department of Ultrasonography, Fudan University Shanghai Cancer Center, Shanghai Medical College, Fudan University, Shanghai, P. R. China

Email: zwenx1123@163.com

Qian Cheng, Institute of Acoustics, School of Physics and Engineering, Tongji University, Shanghai, P. R. China

Email: q.cheng@tongji.edu.cn

**Abstract**

Collagen, a key structural component of the extracellular matrix, undergoes significant remodeling during carcinogenesis. However, the important role of collagen levels in breast cancer diagnostics still lacks effective *in vivo* detection techniques to provide a deeper understanding. This study presents photoacoustic spectral analysis improved by machine learning as a promising non-invasive diagnostic method, focusing on exploring collagen as a salient biomarker. Murine model experiments revealed more profound associations of collagen with other cancer components than in normal tissues. Moreover, an optimal set of feature wavelengths was identified by a genetic algorithm for enhanced diagnostic performance, among which 75% were from collagen-dominated absorption wavebands. Using optimal spectra, the diagnostic algorithm achieved 72% accuracy, 66% sensitivity, and 78% specificity, surpassing full-range spectra by 6%, 4%, and 8%, respectively. The proposed photoacoustic methods examine the feasibility of offering valuable biochemical insights into existing techniques, showing great potential for early-stage cancer detection.

**Abbreviations:** PASA, photoacoustic spectral analysis; PA, photoacoustic; ECM, extracellular matrix; GA, genetic algorithm; ML, machine learning; PAI, photoacoustic imaging; APSD, area of power spectrum density; SVMDA, support vector machine discriminant analysis; KNN, K-nearest neighbor; PLSDA, partial least-squares discriminant analysis; H&E, hematoxylin-eosin; MRI, magnetic resonance imaging.

# 1 INTRODUCTION

Breast cancer is the most common malignancy affecting women, and its incidence has continued to rise by 0.6% per year since 2004, according to Cancer Statistics 2024 [1]. Early-stage breast cancer is curable in approximately 75% of patients [2]. Therefore, early diagnosis is vital for improving survival and prognosis. Mammography is currently the primary method for clinical breast cancer screening [3]. However, it has a risk of radiation exposure and is susceptible to high breast tissue density, which leads to low sensitivity (24‑47%) [4–6]. Ultrasonography is a vastly used adjunct but suffers from operator dependency and insufficient specificity (65-89%) [7,8]. Further, while magnetic resonance imaging (MRI) has been used to evaluate the clinical outcomes of patients, it is restricted by the high costs [9]. Thus, the advent of novel practical tools for providing complementary information and expanding the scope of cancer research is in demand.

The impact of extracellular matrix (ECM) on the behavior of malignant cells has become a developing research area in oncology. Collagen is the most abundant component of ECM that acts as a structural scaffold for tissue homeostasis [10]. Recent studies indicate that collagen degradation and redeposition accompany tumor initiation and progression, thereby modulating ECM to become a favorable place for the interaction, invasion, and metastasis of cancer cells [11]. Literature has reported that the collagen level can be applied as an important biomarker for predicting immunotherapy resistance [12] and median survival of patients [13]. For the characterization of collagen in bio-tissues, current modalities primarily include histopathological staining, second-harmonic generation [14], and mass spectrometry [15]. These methods mainly use *ex vivo* samples, which are invasive and hinder their clinical application in early-stage cancer screening and diagnosis.

Photoacoustic spectral analysis (PASA) is an emerging technique for non-invasive cancer diagnostics since it can capture the composition and physical properties of biomacromolecules based on their unique optical absorption [16]. The photoacoustic (PA) modality is applicable for *in vivo* utility because it can reach deep bio-tissues, not relying on light coherence [17,18].

Ultrasonic signals generated by exposed chromophores experience only one-way propagation before detection, which minimizes the attenuation. The effectiveness of PASA in investigating the microstructures of collagen and lipid for disease evaluation has already been examined in many studies on Crohn's disease [19], cirrhosis or fatty liver disease [20], osteoporosis [21–23], and prostate cancer [24]. Our group proposed a PA parameter "area of power spectrum density (APSD)" to reflect the contents of biomacromolecules, and has successfully used it for identifying myocardial infarction boundary [25], skin tumors [26], and molecular breast cancer subtypes [27]. For breast cancer evaluation nowadays, most studies have mainly focused on PA imaging (PAI) for tumor angiogenesis and hemoglobin oxygen saturation [28]. Studies discussing the potential of PASA to reveal more detailed information about the collagen status of ECM related to carcinogenesis are lacking [29].

On the other hand, various rapidly developing machine learning (ML) methods provide effective ways to extract information from a plethora of collected biomedical data [28–30]. To reveal the intrinsic patterns of physiological processes, unsupervised algorithms like hierarchal clustering, correlation network analysis, and community detection have been widely applied to datasets of genomic sequencing, proteomics, and neuroscience. These methods contribute to understanding the basic mechanism of diseases [30] and the functional connectivity of the brain [31], which also show great potential for *in vivo* PASA to characterize biomacromolecule changes caused by cancer initiation [32]. Furthermore, to enhance the performance of diagnostics, a genetic algorithm (GA) is proposed as a powerful tool for selecting optimal features as inputs for supervised models [33,34]. GA-facilitated small ML models, including K-nearest neighbor (KNN), partial least-squares discriminant analysis (PLSDA), and support vector machine discriminant analysis (SVMDA), were reported applicable for molecular datasets and can achieve higher accuracy with reduced computational cost compared to deep learning models with image inputs [35–37]. Heretofore, the usage of GA in vibrational spectroscopy is restrained to pure optical techniques such as Fourier-transform infrared and Raman spectroscopy [38]. Limited work has been done on PASA to select the optimal set of feature wavelengths for improving cancer theranostics.

This study sheds light on non-invasive *in vivo* breast cancer diagnostics by employing PASA with multiple ML methods to characterize collagen as a biomacromolecule biomarker of ECM remodeling due to carcinogenesis. The murine model was selected for a feasibility study because the xenograft was located at the superficial layer of the mouse body. Furthermore, the collagen in murine malignancy was sufficient to exhibit significant changes from those in normal tissues, as proven by our histopathological studies. A schematic diagram in **Figure 1(a)** explains the relationships of ML algorithms and the logical order of their corresponding research targets. One hundred samples from murine models with a ratio of 1:1 for the malignant and normal were tested to acquire the spectra of a PA parameter APSD using lasers with wavelengths of 1200–1700 nm, where collagen, water, and lipid exhibited significant absorption. In the NIR-II region, hemoglobin, an essential PA agent for vessel imaging, manifested insignificant absorbance [39], causing little bias to the collagen detection based on the PASA technique. Concerning optical absorption spectra provided by previous studies [16,39], the wavelength range demonstrating the most prominent absorption for a certain biomacromolecule among all studied molecules was determined by applying hierarchical clustering to the APSD spectra for *in vivo* detection, which is referred to as the dominated absorption waveband. Then, other unsupervised algorithms were used to analyze the molecular variation between two types of tissues. The collagen content change and its correlation with lipid were discussed. Based on the evidence provided by PASA above, the GA method was utilized on PA spectra to pinpoint an optimized set of wavelengths, termed feature wavelengths, whose APSDs could enhance the performance of breast cancer diagnostics utilizing the most appropriate supervised classifiers.

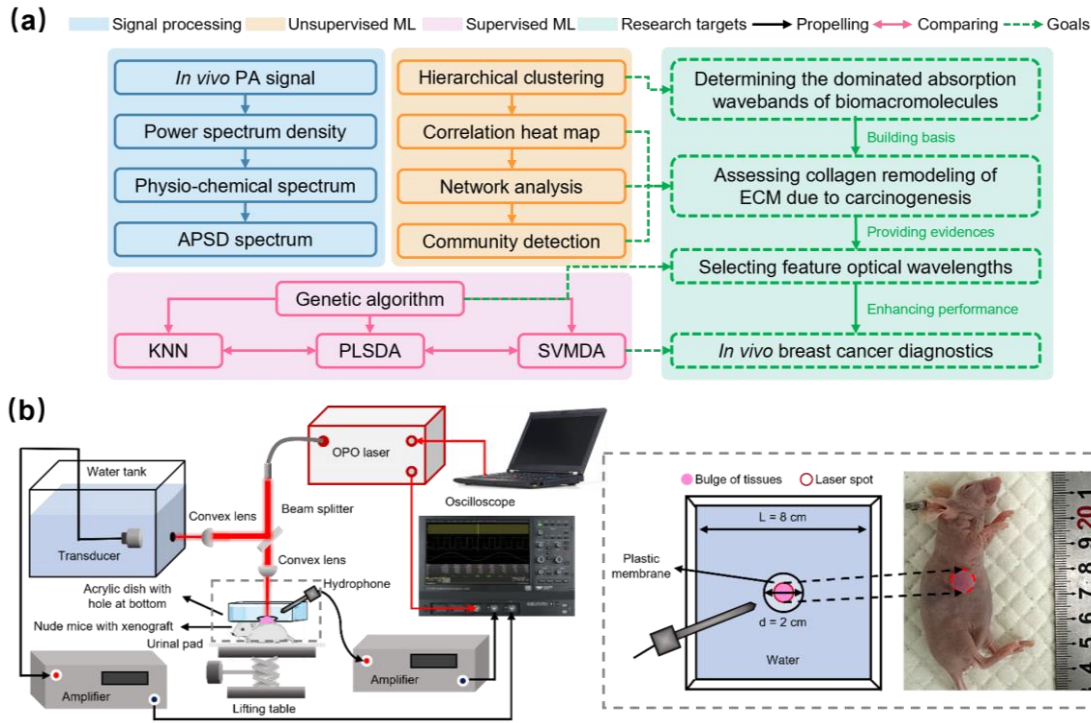

**FIGURE 1**. Overview of the research. (a) Schematic of the approaches and goals in this study. (b) Photoacoustic (PA) experimental system. Details for the ultrasonic coupling devices are shown in the grey dashed box

## 2 MATERIALS AND METHODS

### 2.1 Animal models and experimental protocol

Fifty xenografts were obtained by implanting human breast cancer cells into the mammary glands of nude mice. To evaluate the heterogeneity of breast cancers, we used three distinct cell lines (MCF-7, MCF-10CA1a-HER2, and MDA-MB-231) to establish xenografts representing molecular subtypes of luminal (n=13), HER2 (n=19), and triple negative (n=18), respectively. *In vivo,* PA detection was conducted after the xenografts reached a diameter of 1 cm. Nude mice were anesthetized during the experiments using pentobarbital sodium (1%) via intraperitoneal injection. Normal samples were collected from the opposite tissues of xenografts. The Animal Welfare and Ethics Group of the Department of Laboratory Animal Science, Fudan University, approved the study protocol.

## 2.2 Histopathology and statistical analysis

Nude mice were euthanized after the experiments, and Masson's trichome and hematoxylin-eosin (H&E) staining were performed to assess the collagen and lipid in tissues. To quantify the levels of biomacromolecules, we calculated the relative contents from the cross-sections of xenografts and the normal tissues under PA detection. The relative collagen and lipid contents were defined as the percentage of the positively stained area on the entire section. As illustrated in **Figure S1**, luminal breast cancer had a significantly higher collagen level than do the other two molecular subtypes. Statistical analysis was conducted to compare collagen contents of normal and cancerous samples based on five independently built murine models bearing luminal cancers using t-tests in Prism 9.4.1 software. Collagen and lipid contents of all cancerous samples were also analyzed.

## 2.3 Experimental setup for in vivo PA measurements

The experimental setup (**Figure 1(b)**) comprised laser triggering and ultrasonic signal acquisition parts. A tunable optical parametric oscillator (Phocus Mobile, OPOTEK, Carlsbad, CA, USA) was programmed to emit lasers with wavelengths switching from 1200 nm to 1700 nm at a constant 10 nm interval, a pulse width of 2–5 ns, and a repetition rate of 10 Hz. Collagen, the targeted biomacromolecule of this study, shows a strong light absorption in the above wavebands [39]. The fluence of the laser source is depicted in **Figure S2**. The laser beam was split using a dichroic mirror. One beam with 10% laser energy was focused on the blackbody to produce PA signals for calibrating the laser energy varying with time and wavelength. Here, the blackbody, a piece of black rubber tape pasted on a water tank, had a uniform and stable optical absorption. The ultrasonic signals from the blackbody were collected using a 5 MHz transducer (V302, Immersion Transducers, Olympus Corp., Tokyo, Japan) and amplified by 25 dB using an amplifier (5073PR, Olympus Corp., Tokyo, Japan).

The other beam with 90% laser energy irradiated the sample, with a weak focal spot of 1 cm diameter to cover the entire xenograft, while the optical energy density was controlled to below 25 mJ·cm$^{-2}$ to meet the safety limit specified by the American National Standards Institute for human skin (100 mJ·cm$^{-2}$ for laser wavelengths ranging from 1050 to 1500 nm, and 1 J·cm$^{-2}$ for wavelengths from 1500 to 1700 nm) [40].

The ultrasonic coupling devices were designed to maintain the stability of the experimental system. A water capsule was created using a thin square acrylic dish with a hole of 2 cm diameter at the center of the bottom. The hole was sealed with a thin plastic membrane, and the dish was filled with a shallow 0.4 cm layer of water. During detection, nude mice were raised on a lifting table to let the cancerous or normal tissues slightly press against the capsule. A small bulge was formed for good coupling. The surface distance between the tissues and the water surface was maintained at nearly 1 mm to minimize the water-induced light attenuation. Ultrasonic signals from the xenografts were collected utilizing a needle hydrophone (HNC-1500, ONDA Corp., Sunnyvale, CA) and amplified by 35 dB using an amplifier (5072PR, Olympus Corp., Tokyo, Japan). Signals received a 1 MHz high-pass filter to eliminate low system noise and were stored in an oscilloscope (HDO6000, Teledyne Lecroy, USA) at a sampling rate of 250 MHz. To obtain an adequate signal-to-noise ratio (SNR), 64 average measurements were taken, which required approximately 40 s. Each complete experiment, consisting of 51 wavelengths, took about 34 mins, slightly less than the duration of anesthesia.

**2.4 Multi-wavelength PASA**

The signals were processed using MATLAB software (R2017b, MathWorks, Natick, MA, USA) to construct PA spectra as a raw dataset for further ML analysis in the following steps, Step 1: The PA signals from samples (**Figures 2(a)-(b)**) were divided by the peak-to-peak values of the corresponding blackbody signals; Step 2: The corrected signals were transformed into power spectral density curves (**Figures 2(c)-(d)**) using Welch's method with a moving Hamming window, followed by calibrating under the frequency response of the hydrophone; Step 3: All the density curves for the 51 laser wavelengths constituted the physiochemical spectrogram (**Figures 2(e)-(f)**), where the x-axis represented the optical wavelength, and the y-axis indicated the frequency distribution of ultrasonic signals; Step 4: The PA parameter was calculated by integrating the calibrated power spectral density as follows:

$$APSD_\lambda = \int_{f_0}^{f_1} p(f,\lambda)\mathrm{d}f \tag{1}$$

here, $p(f,\lambda)$ is the calibrated power spectral density, and $f_0 = 1$ MHz and $f_1 = 4$ MHz are the lower and upper boundaries of the frequency range, respectively. The attenuation of most signals exceeded 20 dB beyond the upper limit. $APSD_\lambda$ indicates the power of PA signal at a specific wavelength, which was in direct proportion to the molecular contents in xenografts; Step 5: APSDs from 51 detection wavelengths (1200–1700 nm, $\Delta\lambda = 10$ nm) composed a spectrum (**Figures 2(g)-(h)**) that was unique for each sample. The raw dataset used in this study consisted of APSD spectra of 50 normal samples and 50 cancerous samples.

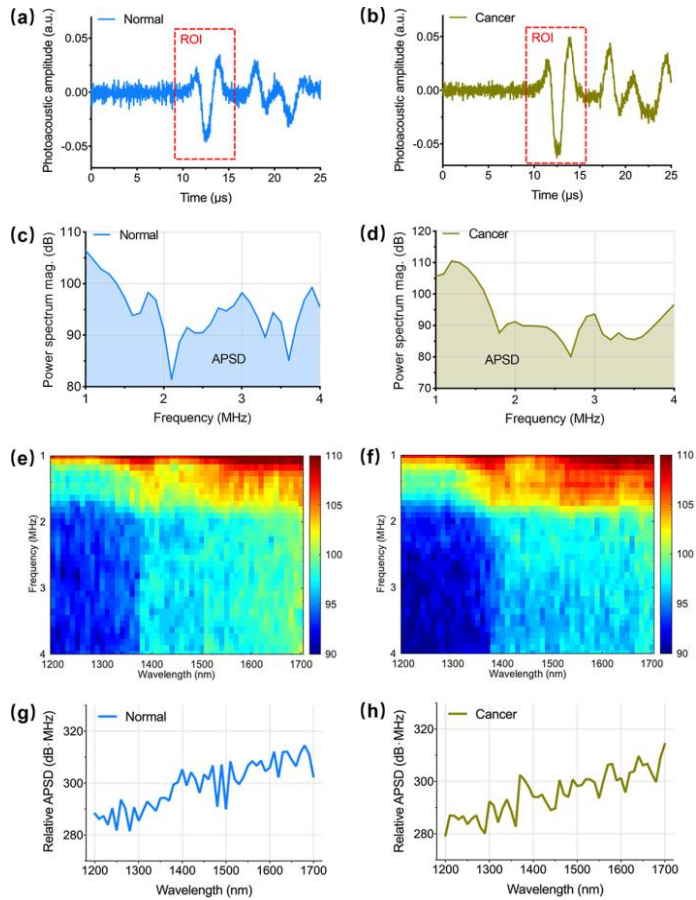

**FIGURE 2.** Acquiring process of the PA spectrum. Representative *in vivo* PA signals of (a) the normal and (b) the cancerous samples. Representative power spectra of (c) the normal and (d) cancerous samples. Mean physio-chemical spectrogram of (e) the normal and (f) cancerous samples. Representative area of power spectrum density (APSD) spectra of (g) the normal and (h) cancerous samples. "mag." refers to a magnitude. "ROI" refers to a region of interest.

## 2.5 Unsupervised machine learning: characterizing carcinogenesis-related biomacromolecule

2.5.1 Hierarchical clustering

The partition of absorption wavebands for studying biomacromolecules in the NIR-II window (1200–1700 nm) was achieved using hierarchical clustering implemented in the R package Pheatmap (version 1.0.12) [41]. Using Euclidean distances, nodes (APSDs) or clusters were linked to the closest node. An agglomerative algorithm was used to determine the communities of the variables (APSDs of each wavelength or sample) based on their similarities. A binary tree illustrated the hierarchical structure of the variables. Variables were standardized using z-scores (mean centering and scaled variance) before analysis.

2.5.2 Correlation heat map

The distribution of APSDs at each wavelength was evaluated for normality using the Shapiro-Wilk test in Prism 9.4.1 software. Then, the associations of sample APSDs at any two wavelengths were computed using Spearman correlation coefficients (SCCs) and visualized using a correlation heat map (51×51 symmetric diagonal matrix). The APSDs of two wavelengths were considered to have sufficient interconnection if their SCC surpassed 0.9. A threshold of 0.9 was applied to the SCCs to distinguish the highly correlated regions.

2.5.3 Wavelength network acquisition and community detection

The correlations of biomacromolecules were further investigated by network analysis of highly correlated regions in the heat map. The elements in these regions were assigned a value of 1, whereas the rest were 0. Consequently, the binary matrix was acquired to build an unweighted and undirected network of wavelengths using the R package Ggraph (version 2.1.0) [42]. The WalkTrap algorithm was used to partition the network into communities using the random walk-based similarity between nodes (wavelengths) and clusters to cut the hierarchical structure [43]. Nodes within the same community were more similar to each other. We calculated the betweenness centrality (BC) of the nodes to evaluate their impact on the network. BC is the frequency of a node found along the shortest path between two other nodes [44].

**2.6 Supervised classification: in vivo breast cancer diagnostics**

2.6.1 GA: feature optical wavelength selection

GA selected feature optical wavelengths for PASA-based breast cancer diagnostics to enhance model performance while reducing required wavelengths [45]. Step 1: A population of 64 individuals with random subsets of wavelengths was generated. Root mean square error for cross-validation (RMSECV) was calculated using multiple linear regression (MLR) between individual variables and sample response variables (1 for cancerous, 0 for normal). Step 2: Poor-performing individuals (RMSECV > median) were excluded; the remaining individuals exchanged random variables (double crossover) to replenish the population. Step 3: New populations were re-evaluated, discarded, mutated, and bred until steps 1 and 2 were repeated reaching 50 generations. These three steps were performed with 100 replicate runs to build a collection of optimal models. The variables resulting in the lowest RMSECV were selected as feature wavelengths.

2.6.2 KNN

For a more comprehensive consideration, three supervised classifiers applicable to small datasets were tested to compare their performances on GA-optimized PASA for breast cancer diagnostics. All approaches were implemented in MATLAB software (R2017b, MathWorks, Natick, MA, USA) with the PLS Toolbox v.9.3 (Eigenvector Research Inc., Wenatchee, WA, USA). KNN is a distance-related model that assigns the class of an unknown sample by the vote of a small group of its nearest neighbors [45]. K is commonly chosen as an odd number ranging from 3 to 9. The closeness of the samples was calculated using the Euclidean distance.

2.6.3 PLSDA

PLSDA is a linear classifier developed from the partial least-squares regression [46]. Linear regression was performed between the latent variables (LVs) extracted from the predictors (APSD spectra) and the response variables, set as the same as GA. Several independent LVs with eigenvalues > 1 were extracted to describe the maximum amount of data variance. Samples whose statistics Q residuals and Hotelling's $T^2$ exceeded the 99% confidence level were excluded from the dataset as outliers until the final model was constructed [47]. A probabilistic threshold was calculated to identify the cancerous samples, assuming a Gaussian distribution of the regression values.

2.6.4 SVMDA

This study uses SVMDA as a nonlinear classifier, adjusting a hyperplane with an optimal margin to classify samples projected into high-dimensional spaces [48]. A radial basis function (RBF) with a Gaussian kernel type was used to transform the PA dataset. The shape of the hyperplane was modified by the gamma ($\gamma$) to alter the RBF. The width of the margin was penalized by the cost (c) to constrain the support vectors lying within it. A search over grids from $10^{-6}$ to $10^9$ uniformly spaced in the log was applied to determine the appropriate value for $\gamma$ and c, minimizing the misclassification fraction in the testing set [47].

2.6.5 Model validation and evaluation

The reliability of supervised ML models was ensured using ten-fold cross-validation [49]. The samples were divided into ten equal-sized subsets. Each subset was used as a testing set once to verify the classifiers built employing the remaining subsets (calibration sets). The mean accuracies, specificities, and sensitivities were calculated for the calibration and the testing sets. Details regarding the sample allocation are listed in **Table 1**. The confusion matrices, receiver operating characteristic curve (ROC), and the area under the curve (AUC) were also considered to evaluate the performance of the model.

**TABLE 1:** Sample size for the supervised machine learning (ML) models

| Label  | Calibration set | Testing set | Total |
|--------|-----------------|-------------|-------|
| Normal | 45              | 5           | 50    |
| Cancer | 45              | 5           | 50    |
| Total  | 90              | 10          | 100   |

## 3 RESULTS

### 3.1 Histopathological analysis of normal and cancerous tissues

Masson's trichrome staining was performed on both normal and cancerous tissues to assess the relative collagen content. As depicted in the statistical results in **Figure 3(a)**, normal samples owned a significantly elevated collagen level. Representative staining images are exhibited in **Figure 3(b)**. Observations revealed that collagen fibers formed a loose and regular network in normal samples while they became disordered in cancerous samples. This finding was also reported in other literature [50–52]. The histopathological results proved that collagen underwent significant remodeling in the murine xenografts, providing a solid basis for PASA for cancer diagnostics.

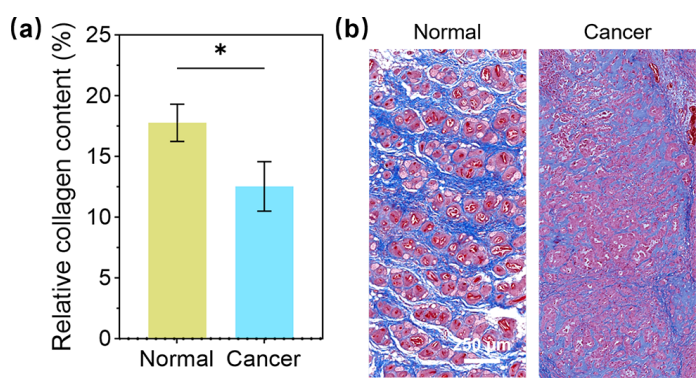

**FIGURE 3.** Characterizing collagen content of normal and cancerous tissues through Masson's trichome staining. (a) Statistical analysis (n=10, *$p < 0.05$). (b) Representative staining images. The statistical analysis was assessed using the t-test.

### 3.2 Determining the dominated absorption wavebands of biomacromolecules

Among the optical absorption coefficients of the three biomacromolecules shown in **Figure 4(a)**, collagen has a gradually increasing absorbance in the NIR-II window (1200–1700 nm) [21,32]. Lipid has a growing absorbance at wavelengths beyond 1650 nm. To eliminate the spectra bias caused by testing concentrations, the absorption spectra were normalized as revealed in **Figure 4(b).** The tape plot in **Figure 4(c)** interprets the APSD spectra of all the tested samples and their standard deviations (SDs). Compared to the normal samples, the cancerous samples exhibited similar profiles but had approximate APSDs in 1410-1520 nm or lower APSDs in 1200-1400 nm and 1530-1700 nm, which was consistent with our histopathological findings that cancerous samples had relatively lower collagen contents than normal samples. Moreover, the cancerous group showcased a more extensive variation of APSDs; this could be interpreted by the fact that ECM deteriorated with cancer initiation. The inter-cancer heterogeneity of biomacromolecules increased due to the malignant differentiation [53].

The hierarchical clustering heat map of the APSD dataset is shown in **Figure 4(d)**. It was processed in two directions: wavelengths and samples. The binary tree of laser wavelengths drawn along the rows was partitioned into three distinct segments. This division result determined the dominant absorption wavebands of collagen, water, and lipid for the *in vivo* PA study. A schematic axis (**Figure 4(e)**) clarifies the absorption wavebands: collagen-dominated (1200-1400 nm and 1530-1630 nm), lipid-dominated (1640-1700 nm), and water-dominated (1410-1520 nm). The classification, consistent with the normalized optical absorption depicted in **Figure 4(b)**, provides a basis for *in vivo* detecting biomacromolecules under specific wavebands. Two separate wavebands, 1200-1400 nm and 1530-1630 nm, were considered for collagen because the literature reported that both 1370 nm [54] and 1550 nm [23] lasers were exploited for collagen measurements. In the axis of **Figure 4(e)**, "Collagen 1" and "Collagen 2" symbolize two detection windows for collagen, unrelated to collagen subtypes. The review [55] indicates that no research has yet validated variations in PA signals across collagen subtypes. On the other hand, the binary tree of samples was drawn along the columns. The clustering results appeared disordered for two groups of samples; this suggested that more elaborate ML techniques were needed to unravel the underlying patterns of molecular changes induced by carcinogenesis.

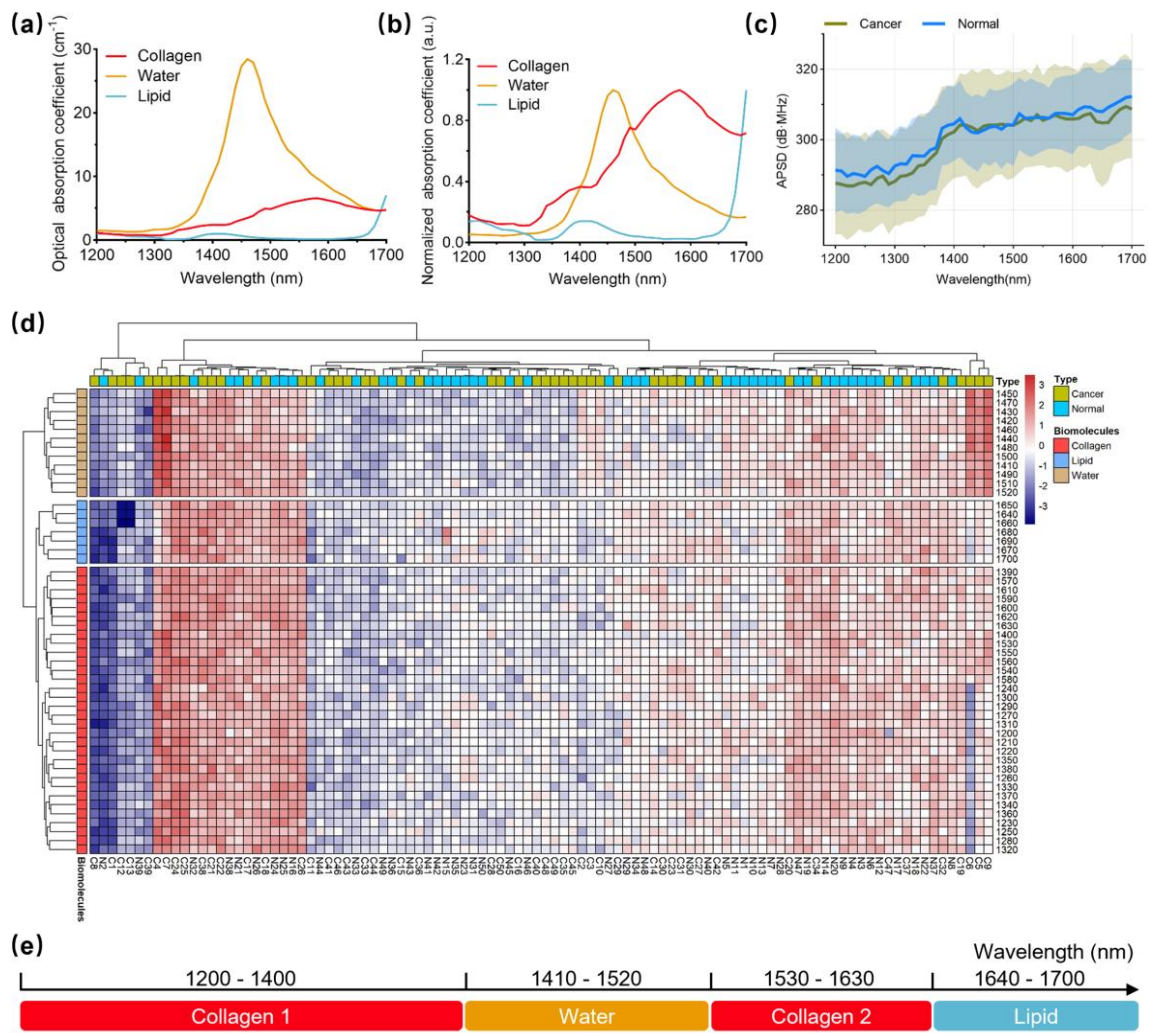

**FIGURE 4.** Determining biomacromolecule-dominated absorption wavebands. (a) Optical absorption spectra of key biomacromolecules in the NIR-II window (1200–1700 nm): collagen, lipid, and water. (b) Normalized optical absorption coefficients. (c) Relative APSD spectra for all the samples (n=100), with solid lines indicating mean values, and shaded areas representing standard deviations (SDs). (d) Hierarchical clustering of APSDs across different laser wavelengths (rows) and samples (columns), with samples labeled as N (normal, n=50) and C (cancerous, n=50). (e) Partitioning of the dominated absorption wavebands for the three

biomacromolecules, with two collagen-dominated wavebands labeled as Collagen 1 and Collagen 2.

**3.3 Assessing collagen remodeling of ECM due to carcinogenesis**

The SCCs of the APSD spectra for all the wavelengths are shown as heat maps (**Figures 5(a)-(b)**), serving as a foundation for characterizing the correlation of biomacromolecule content in tissues. Both maps of normal and cancerous samples revealed all SCCs exceeding 0.4. Intriguingly, the cancerous samples represented an elevated correlation of APSDs. The APSD spectra from the collagen-dominated (1200‑1400 nm and 1530‑1630 nm) and the lipid-dominated absorption wavebands (1640‑1700 nm) shared more intense similarities than the water-dominated absorption wavebands (1410‑1520 nm). These results embodied the physiological environment change due to the formation of breast cancer.

The most correlated regions with SCCs > 0.9 in the heat map of cancerous samples are drawn in **Figure 5(c)**. Mutual relations of collagen and lipid appeared in the regions outside the blocks restricted by specific biomacromolecules. This cross-impact was also observed in our previous study on *ex vivo* prostate-tissue specimens [56]. The wavelength networks of the cancerous samples are exhibited in **Figure 5(d)**. Nodes were arranged clockwise first by the size of communities and then the number of edges between them and other nodes. Communities were specified by the biomacromolecules whose absorption-dominated wavebands occupied the majority of wavelengths. The radial layout of the network and communities was contrived to deliberate on the molecule-dependent impact. Three significant communities were discovered for collagen and water. The most prominent communities consisted of wavelengths chiefly from the collagen-dominated absorption waveband. Wavelengths in the lipid-dominated absorption waveband but assigned to the collagen-dominated community are highlighted in blue boxes. These traits indicated that collagen content had profound linkages with lipid in the tumor ECM.

The normalized BCs of nodes are represented in **Figure 5(e)** to assess the positions of wavelengths within the network. A wavelength owning higher BC indicated that its APSD had more direct or indirect connections with APSDs of the other wavelengths. The prominent peaks, appearing at 1590 nm, 1520 nm, and 1620 nm, indicated the significant influence of collagen-dominated absorption wavebands on the correlation network, suggesting that evident collagen remodeling of ECM accompanied carcinogenesis and might coordinate the behaviors of lipid.

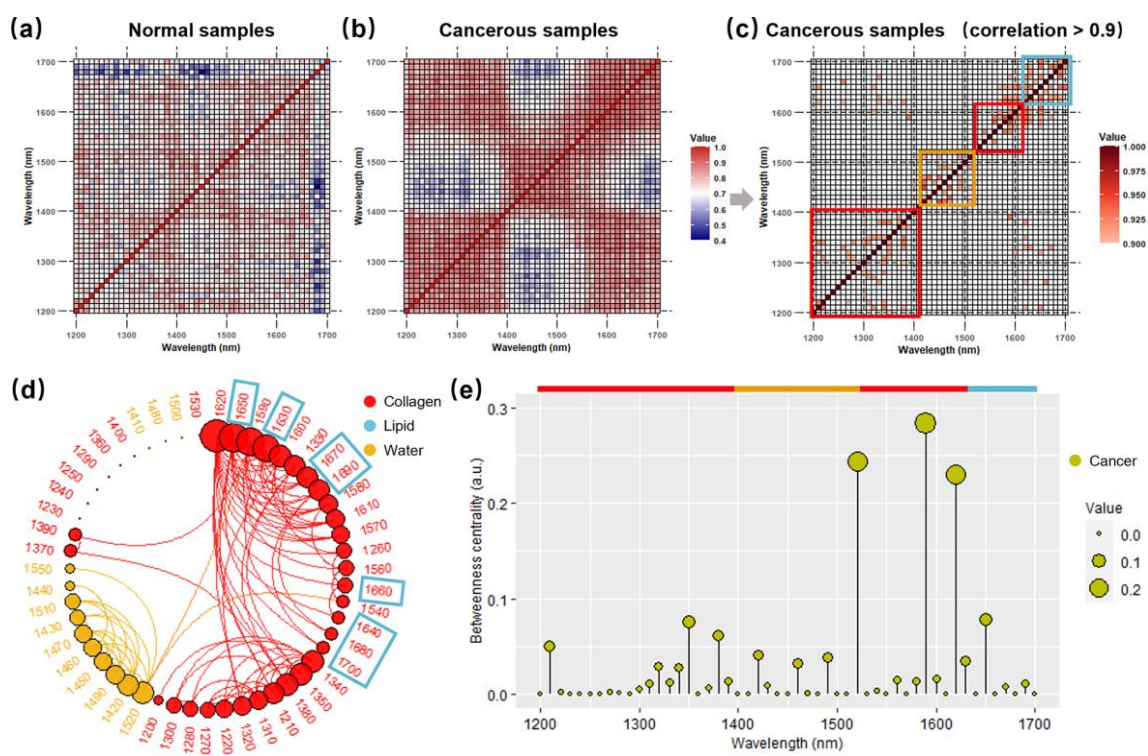

**FIGURE 5.** Assessing the cancer-related biomacromolecule change of the extracellular matrix. (a) Correlation heat map for normal samples (n=50). (b) Correlation heat map for cancerous samples (n=50). (c) Correlation heat map for cancerous samples after applying a threshold of 0.9. (d) Laser wavelength network of the cancerous samples displayed in a radial layout. (e) Lollipop chart showing the betweenness centralities (BCs) of nodes within the network. The colors of the boxes, nodes, and bars represent the dominant absorption wavelengths of collagen (red), lipid (blue), and water (gold).

**3.4 In vivo breast cancer diagnostics based on PASA**

3.4.1 GA-based feature optical wavelength selection

The GA technique built 158 MLR models through 100 runs, 50 generations per run. **Figure 6(a)** depicts the RMSECVs of all the models varying with wavelengths by rows, arranging in decreasing values from the top to the bottom. The inclusion frequencies of the wavelengths within the 158 models are shown in **Figure 6(b)**. The wavelengths with frequencies greater than 0.5 are labeled above the bars. Overall, 83% of these frequently employed wavelengths were from collagen-dominated absorption wavebands. **Figure 6(c)** interprets the RMSECVs of all the models depending on the number of included wavelengths, varying from 4 to 10. The best regression model owning the lowest RMSECV was established by APSDs from 8 wavelengths: 1280, 1300, 1330, 1360, 1400, 1430, 1510, and 1540 nm. These wavelengths were selected as the feature optical wavelengths for breast cancer diagnostics, among which wavelengths belonging to the collagen-, water-, and lipid-dominated absorption wavebands accounted for 75%, 25%, and 0%, respectively.

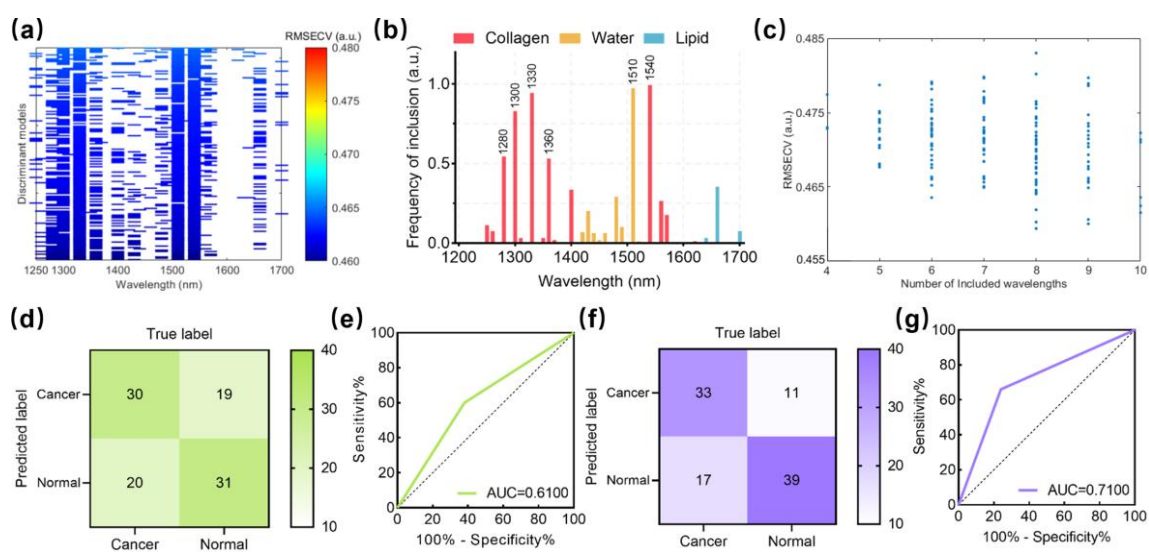

**FIGURE 6.** Diagnostics of breast cancers using genetic algorithm (GA)-facilitated machine learning classifiers via photoacoustic spectral analysis. (a) All 158 iterative models were obtained via 100 replicate runs of GA. (b) Frequencies for wavelength inclusion in all the GA models. The wavelengths included in more than half of the models are labeled. (c) Root mean

square error for cross-validation (RMSECV) of all the GA models depends on the number of wavelengths. (d)-(e) Confusion matrix, receiver operating characteristic curve (ROC), and area under the curve (AUC) for the K-nearest neighbor model based on full-range wavelength spectra. (f)-(g) confusion matrix, ROC, and AUC for the support vector machine discriminant analysis model based on feature optical wavelength spectra.

3.4.2 Enhanced discriminant models utilizing feature wavelength spectra

Although full-range wavelength PA spectra (1200–1700 nm, $\Delta\lambda = 10$ nm) contained complete information on biomacromolecules, some redundant features might prevent ML classifiers from projecting the normal and cancerous samples into proper data spaces and deciding a precise partition for identification. The performances of supervised ML models for breast cancer diagnostics are summarized in **Table 2**. While using full-range wavelength spectra, the linear PLSDA model and the nonlinear SVMDA model led to low sensitivity or specificity on the calibration set. Compared to these two algorithms, the distance-based KNN model had relatively balanced accuracies for the negative and positive samples on the cross-validation sets but only demonstrated a moderate diagnostics capability. The confusion matrix and the ROC with AUC for the testing set of KNN are displayed in **Figures 6(d)-(e)**.

Furthermore, while using feature wavelength (8 wavelengths obtained from GA) APSD spectra, all the ML models underwent a performance elevation. The PLSDA model had more evident progress than the KNN model. However, its sensitivity deteriorated sharply compared to specificity. The SVMDA model revealed the best capability among the three ML algorithms. **Figures 6(f)-(g)** exhibit the confusion matrix and ROC for the testing set of SVMDA. The results interpreted that GA-selected feature optical wavelength spectra with a nonlinear SVMDA model could improve the accuracy, sensitivity, and specificity of breast cancer diagnostics via PASA by 6%, 4%, and 8% in comparison with the full-range wavelength spectra.

**TABLE 2:** Comparison of the performances of supervised ML models

|  |  | Full-range wavelength APSD spectra | | | Feature wavelength APSD spectra | | |
|---|---|---|---|---|---|---|---|
|  |  | KNN | PLSDA | SVMDA | KNN | PLSDA | SVMDA |
| Calibration | Accuracy | 57% | 60% | 36% | 61% | 71% | 76% |
|  | Sensitivity | 56% | 36% | 62% | 56% | 60% | 76% |
|  | Specificity | 58% | 80% | 10% | 66% | 82% | 76% |
| Testing | Accuracy | 61% | 63% | 66% | 60% | 65% | 72% |
|  | Sensitivity | 60% | 48% | 62% | 56% | 54% | 66% |
|  | Specificity | 62% | 76% | 70% | 64% | 76% | 78% |
| Parameters |  | K = 3 | LVs = 1 | c = $10^{-1.0}$, γ = $10^{-1.5}$ | K = 3 | LVs = 2 | c = $10^{3.0}$, γ = $10^{-3.0}$ |

## 4 DISCUSSION

Clinical breast cancer diagnostics currently rely primarily on mammography and ultrasonography. While these modalities are extensively used, they fail to provide information beyond tissue morphology. PASA, which is sensitive to biochemical changes, offers a deeper molecular understanding of ECM and shows great potential for early cancer diagnostics. Collagen, a major fibrous protein in the ECM, has been shown to exhibit significant remodeling of content and arrangement during cancer initiation and growth, creating a fertile ECM that facilitates cancer cell proliferation and invasion [57]. Therefore, collagen is proposed as a biomacromolecule biomarker for cancer theranostics.

However, the application of PASA in cancer research poses great challenges. The complex compositions of cancer tissues make it difficult to directly interpret information about specific biomacromolecules. PASA, powered by multiple ML methods, offers a comprehensive molecular insight into breast cancer without requiring explicit knowledge of tissue composition. This modality aims to improve cancer diagnostics and reduce detection costs. Additionally, unlike mammography, the PA technique is not strongly influenced by high breast density [8], benefiting populations at increased risk. The above benefits of PASA with ML can expand the scope of traditional clinical modalities.

The current study analyzed the feasibility of improving *in vivo* PASA using ML to explore collagen change as a prominent biomarker for breast cancer diagnostics. Histopathological studies demonstrated a significant difference in collagen contents between the murine normal and cancerous tissues. The spectra of APSD, a PA parameter reflecting molecular contents, were investigated. First, the joint use of unsupervised ML algorithms was applied to mine out the characteristics of biomacromolecules. The dominated absorption wavebands of collagen (1200‑1400 nm and 1530‑1630 nm), water (1410‑1520 nm), and lipid (1640‑1700 nm) were confirmed as a basis for *in vivo* PA studies by hierarchical clustering. The correlation heat map revealed that collagen content varied significantly between the normal and cancerous tissues. Collagen and lipid had a high correlation of content in the tumor ECM where the bio-tissue heterogeneity was enhanced. The wavelengths in the two largest communities of network analysis results were mainly from collagen-dominated absorption wavebands, with wavelengths owning the maximal BCs also within this waveband, suggesting the synergistic impact of collagen remodeling on other ECM biomacromolecules due to carcinogenesis. Compared to algorithms used for unmixing PA spectra, the performance of proposed PASA methods is not strongly influenced by the optical absorption spectra selected for biomacromolecules. Due to the substantial variation in the physiological environment, determining precise absorption spectra to use as a benchmark for spectral unmixing remains challenging.

Based on the cancer-induced variation of bio-condition validated above, the feature optical wavelengths for breast cancer diagnostics were selected by GA to enhance the performance of supervised ML classifiers. Eight wavelengths were obtained from the best of 158 GA models: 1280, 1300, 1330, 1360, 1400, 1430, 1510, and 1540 nm, in which collagen-dominated absorption wavebands occupied almost the entirety. Previous research has demonstrated lipid absorption peak at 1200–1250 nm [26,58]. Since this waveband did not exist in the feature wavelengths or possessed minimal BCs, it did not impact on the findings of this study. Furthermore, The SVMDA model exploiting feature wavelength spectra, reducing by 85% of the scanning wavelength number, achieved 72% accuracy, 66% sensitivity, and 78% specificity of diagnostics, surpassing those reached by the distance-based KNN model using full-range wavelength spectra (66%, 62%, and 70%, respectively). This improvement demonstrates the advantages of feature wavelength PA spectra over the sophisticated full-range wavelength dataset. Furthermore, GA proved to be a more potent ML technique for data dimension reduction than other algorithms such as principal component analysis. **Figure S3** illustrates that primary LVs (eigenvalues > 1) fail to differentiate between the two sample groups. In general, the GA-optimized spectra could capture the most crucial molecular information reflecting the formation of cancerous lesions. Since the data redundancy was minimized in feature wavelength spectra, the SVMDA could take full advantage of its nonlinear kernel to partition samples in high dimensional spaces.

The PASA interpreted here is a preclinical feasibility study with certain limitations. First, the accuracy of diagnostics using supervised classification models can still be enhanced. This preliminary study used raw PA signals for spectral analysis, neglecting the light attenuation in tissues. Signal correction to compensate for optical fluence, which decreases with depths, shows promises of improving accuracy [59], but a well-established 3D model for detailed cancer composition is required for simulations. We are dedicated to building a collagen distribution model based on Masson's trichome staining images, which will be reported in our future papers. Second, the histopathological studies based on biopsy slides had limited ability to reflect the condition of entire tissues detected by the PA method which was a 3D detection technique. This discrepancy could hinder histopathological analysis from providing a precise evaluation of PA findings. In future works, we plan to increase the number of slides per sample or test other molecular analysis modalities like mass spectrometry. Third, other biomacromolecules such as lipid were not emphasized in this paper. The murine models established here were under basic feeding standards, owning quite lower lipid contents assessed by histopathology compared to collagen as displayed in **Figure S4(a)-(b)**. This might reduce the accuracy of lipid content statistics and prevent embodying the role of lipid in carcinogenesis. We plan to build lipid-rich murine models by incrementally increasing fat in their diet in our future work.

Despite these limitations, this study provides effective methods for *in vivo* characterizing collagen content as an important biomacromolecule biomarker to achieve breast cancer diagnostics based on murine models using improved PASA aided by ML. For future generalization to human subjects, considerable effort is required. As human mammary tissues may differ in collagen compositions and lipid levels, the proposed methods can contribute to re-evaluating the relationship between these two important biomacromolecules and determining the feature wavelengths for cancer diagnostics. To translate the technique into clinical trials, PASA can be readily integrated into existing ultrasound imaging and PAI systems, realizing multimodal measurements. The primary technical challenges will arise from the thick tissues obscuring the human breast cancer lesions. Light attenuation in the covering tissues is likely to be strong and wavelength-dependent, resulting in spectral coloring and reduced SNR. To alleviate these problems, ultrasonic localization will be engaged to isolate the PA signals from the ROI [60], along with model-based fluence correction techniques to augment the desired signals. Fusing PA spectra with other data types, such as multi-omics results, to develop a holistic view of cancer development using powerful ML algorithms will also be a promising research orientation in the future.

# 5  CONCLUSION

This study tested the feasibility of PASA powered by ML for in vivo measurements of collagen as a biomacromolecule biomarker of ECM remodeling to realize breast cancer diagnostics, focusing on murine models. A spectral parameter APSD was calculated to semi-quantify molecular contents. Unsupervised ML methods, including hierarchical clustering, correlation heat map, and network analysis, were employed to fully investigate the characteristics of collagen and its correlation with lipid. The GA was utilized to mine out an optimal set of optical wavelengths whose PA spectra contained the most essential information for breast cancer diagnostics. We found that the heterogeneity of ECM was enhanced due to carcinogenesis, and collagen had intensive correlations with lipid in cancerous environments. The GA-selected feature wavelength spectra achieved 72% accuracy, 66% sensitivity, and 78% specificity based on the SVMDA model, exceeding full-range spectra with the KNN model by 6%, 4%, and 8% while saving 85% of the scanning wavelengths. 75% of the feature wavelengths were from the collagen-dominated absorption waveband. Histological analysis confirmed a significant difference between the collagen contents of normal and cancerous tissues in murine models. The proposed methods provide complementary biochemical information on bio-tissues to current clinical modalities and have great potential to advance breast cancer theranostics.

**ACKNOWLEDGMENTS**

This work was supported by the National Natural Science Foundation of China (grant numbers 12034015 and 62088101), the Natural Science Foundation of Shanghai (grant number 23ZR1412400), the Program of Shanghai Academic Research Leader (grant number 21XD1403600), and the Shanghai Municipal Science and Technology Major Project (grant number 2021SHZDZX0100).We would like to thank Editage (www.editage.cn) for English language editing.

## AUTHOR CONTRIBUTIONS

J.L. was involved in conceptualization, methodology, software, data curation, formal analysis, writing—original draft preparation. L.B. was involved in investigation, resources. Y.C. was involved in methodology, investigation. J.C. was involved in visualization. J.Z. was involved in supervision. W.Z. was involved in resources, funding acquisition. Q.C. was involved in funding acquisition, writing—review and editing.

## CONFLICT OF INTEREST

The authors declare no financial or commercial conflict of interest.

## DATA AVAILABILITY STATEMENT

The data that support the findings of this study are available from the corresponding author upon reasonable request.

**SUPPORTING INFORMATION**

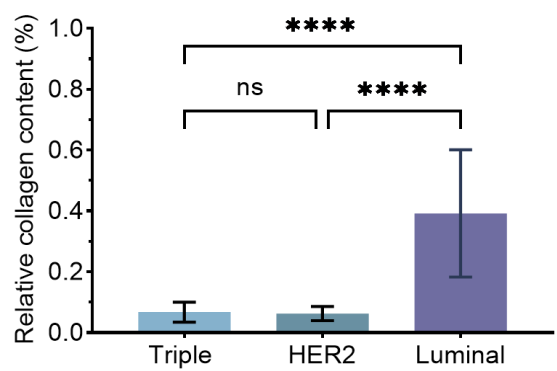

**Figure S1**. Characterizing collagen content of the cancerous tissues of three molecular breast cancer subtypes: triple negative (n=18), HER2 (n=19), and luminal (n=13). The statistical analysis was assessed using the Kruskal-Wallis test and Dunn's multiple comparison tests.

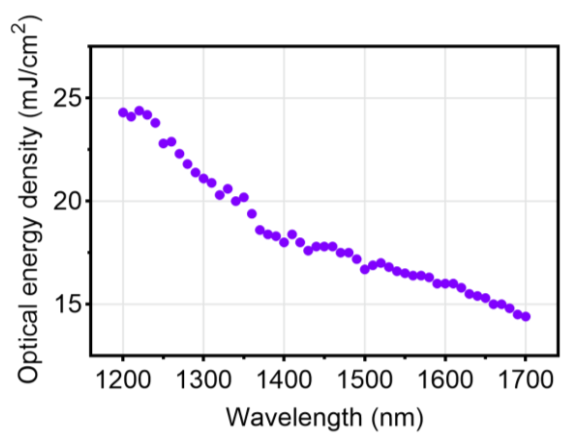

**Figure S2.** Laser fluence for photoacoustic detection over the wavelengths of 1200–1700 nm.

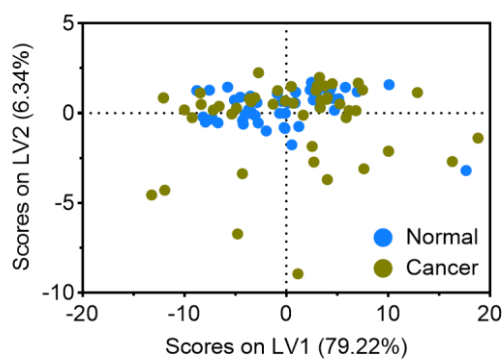

**Figure S3.** Principal component analysis scores of all samples (n=100) on the two extracted latent variables (LVs) with eigenvalues greater than one.

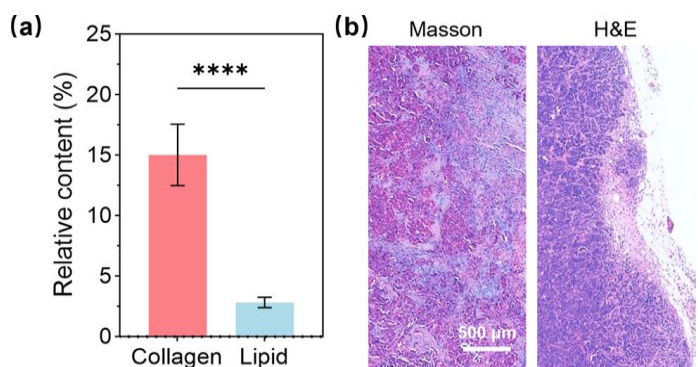

**Figure S4.** Characterizing collagen and lipid content in cancerous tissues through Masson's trichome and H&E staining. (a) Statistical analysis (n=50, ****$p < 0.0001$). (b) Representative staining images. The statistical analysis was assessed using the t-test.